\documentclass[%
twocolumn,
preprintnumbers,
 amsmath,amssymb,
 aps,
 pra,
 notitlepage,
]{revtex4-2}

\usepackage{graphicx}
\usepackage{dcolumn}
\usepackage{bm}
\usepackage{hyperref}

\begin{document}

\preprint{arxiv}

\title{Field configurations for field-free RF trap networks}

\author{Janus Wesenberg}
\email{jhw@kvantify.com}
\affiliation{%
 Kvantify
}

\date{May 1, 2026}

\begin{abstract}
We develop a constructive framework for designing radio-frequency (RF) trap networks from planar data and show that non-smooth field-free guide lines are possible in such networks.
Given analytic Cauchy data on a symmetry plane, namely the potential and its normal derivative, Laplace's equation determines a local three-dimensional continuation.
The odd subclass of this harmonic extension maps an arbitrary analytic generating function $P(x,y)$ to a harmonic potential whose in-plane radio-frequency null set is exactly $P(x,y)=0$.
This yields explicit field-free guide networks beyond smooth straight-line intersections, including cusp guides, cotangential contacts, and periodic lattices.
We further derive Fourier-space formulas for periodic extensions and present square-lattice network families with tunable local crossing angle and rounded connectivity.
These results provide a compact parametrization for the design space for quantum charge-coupled device (QCCD) architectures.

\end{abstract}

\maketitle

RF trap networks in which trapped ions can be moved between trap zones in a fully connected network is a central part of the the QCCD architecture \cite{kielpinski2002:architecture_large_scale}, which is a leading contender for a scalable error-corrected quantum computing architecture.
The RF trap networks used in QIP rely on a combination of control fields that change slowly compared to trap dynamics, and a constant ponderomotive potential provided by an RF field that oscillates at a frequency much faster than trap dynamics.
Because the control potential is provided by a static field governed by Laplace's law, it follows that any net confinement must be provided by the ponderomotive potential.
In Ref.~\cite{wesenberg2009:ideal_intersections_radiofrequency}, we have previously discussed the design considerations for intersections in such trap networks, and identified all possible ideal intersections, i.e. RF field configurations corresponding to a ponderomotive potential which would fully confine ions to a set of field-free guide lines with a first order transverse restoring force at all locations, under the explicit assumption that all such guide-lines would be smooth curves.
So far, no field-free intersections have been demonstrated, but controlled ion transport through non-ideal intersections has been demonstrated by "surfing" the ponderomotive potential at a path with sufficient total confinement, so that a suitable local effective potential can be engineered with the control fields \cite{wright2013:reliable_transport_microfabricated}.
Despite these successes, it would be advantageous to use ideal or field-free intersections if such could be produced: not only would such intersections allow for simpler controls and less RF-heating, but they might enable ballistic transports which could lift some of the topological restrictions on connectivity in cycled trap network.

In this work we show that guide-lines of zero RF-field need not be smooth, and that the catalog of ideal intersections constructed in Ref.~\cite{wesenberg2009:ideal_intersections_radiofrequency} may consequently be incomplete.
In doing so, we develop a method for computing field configurations for a large class of planar trap networks and give an example of a full field configuration for a trap network with hexapole intersections.

This article is structured as follows:
In the first section, we introduce the harmonic extension of planar Cauchy boundary conditions as a strong theoretical tool for working with coplanar trap networks.
As a special case, the following section shows how the zeros of an arbitrary analytical function in the plane can be extended to a field-free traps via the odd part of this harmonic extension.
As an immediate consequence, we show that field-free lines may have cusps, so that the characterization of field-free intersections presented in \cite{wesenberg2009:ideal_intersections_radiofrequency}, may have not fully explored the space of possible intersections.
In the following section, we extend the odd harmonic extension to periodic structures and present a family of trap networks with rectangular grid connectivity.
Lastly, we discuss the possible next steps, including the viability of creating the field configurations we have identified, and the application of the CK-method to non-coplanar networks where we have preliminary indications that non-through intersections are possible. Here we also emphasize that our construction is not exhaustive and that qualitatively different trap network families may be possible.

\section{Harmonic extension of planar Cauchy boundary conditions}

For the typical dimensions (<mm) and frequencies (>MHz) used in QIP RF traps, we may consider the RF field as a quasi-static field, $\bm{E}_{\text{RF}}(\bm{r}, t) = -\bm{\nabla} \Phi(\bm{r}) \cos(\Omega t)$.
The associated ponderomotive potential is
\begin{equation}
    U_{\text{PP}}(\bm{r}) = \frac{Q^2}{4 M \Omega^2} \lvert \bm{\nabla} \Phi(\bm{r}) \rvert^2,
\end{equation}
where $Q$ and $M$ are the charge and mass of the trapped particle.

To parametrize the possible field configurations in a source-free region containing the plane $z=0$, we will consider a harmonic extension of the Cauchy boundary conditions (i.e.~potential and normal derivative) at the plane.
In order to do so, we expand $\Phi$ in powers of $z$ as
\begin{align}
\label{eq:phi_n}
\Phi(x,y,z)=\sum_{n=0}^{\infty}\frac{z^n}{n!}\phi_n(x,y)
\end{align}
with $\phi_n=\partial_z^n\Phi(x,y,0)$.
Defining the in-plane Laplacian $\Delta_{xy}\equiv \partial_x^2+\partial_y^2$, it follows from Laplace's equation $\nabla^2\Phi=\partial_x^2\Phi+\partial_y^2\Phi+\partial_z^2\Phi=0$ that $\partial_z^2\Phi=-\Delta_{xy}\Phi$.
Expanding these expressions and equating powers in $z$, we find that 
\begin{align}
\phi_{n+2}=-\Delta_{xy}\phi_n,
\end{align}
defining two independent recursion relations between the even and odd $\phi_n$.

It follows that we may express the full potential uniquely in terms of $\phi_0$ and $\phi_1$ as
\begin{align}
\label{eq:CK}
\Phi=\Phi_{\rm even}[\phi_0]+\Phi_{\rm odd}[\phi_1],
\end{align}
with
\begin{align}
\Phi_{\rm even}[\phi_0]&=\sum_{m=0}^{\infty}\frac{(-1)^m z^{2m}}{(2m)!}\Delta_{xy}^m \phi_0,\\
\Phi_{\rm odd}[\phi_1]&=\sum_{m=0}^{\infty}\frac{(-1)^m z^{2m+1}}{(2m+1)!}\Delta_{xy}^m \phi_1.
\end{align}

Together, $\phi_0(x,y) = \Phi(x,y,0)$, the potential in the plane, and $\phi_1(x,y) = \partial_z \Phi(x,y,0)$, the normal derivative of the potential, specify the Cauchy boundary conditions for the electric potential in the source-free region, and \eqref{eq:CK} demonstrates how these boundary conditions can be extended to the surrounding source-free region.

\section{Planar traps from odd harmonic extension}

\begin{figure}[t]
\centering
\includegraphics[width=1.0\linewidth]{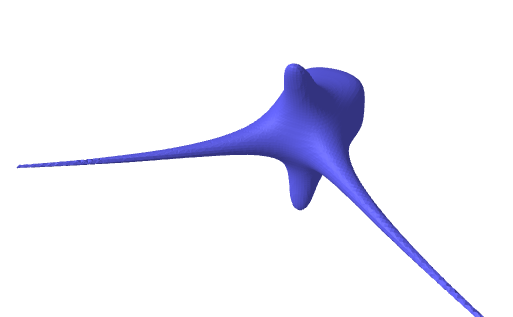}
\includegraphics[width=0.7\linewidth]{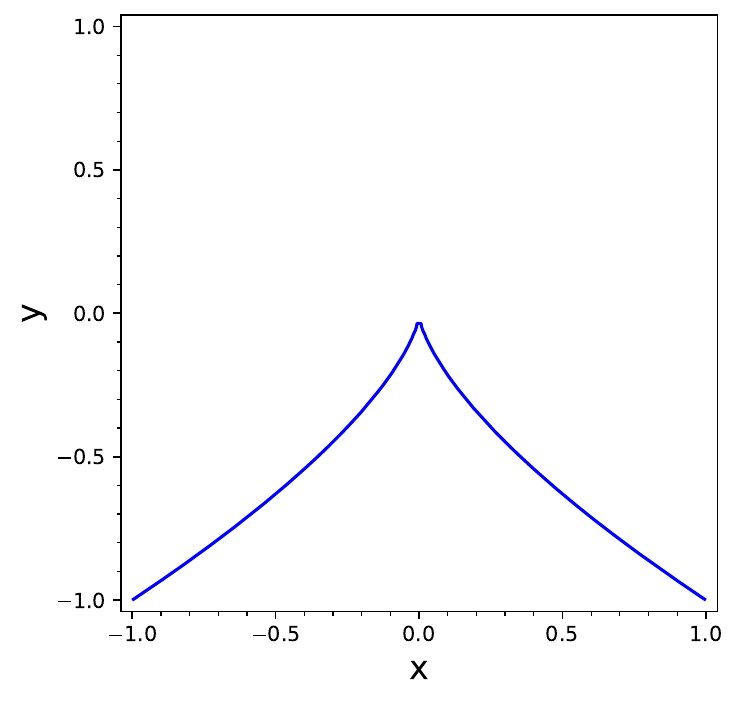}
\caption{
\label{fig:cusp}
Null lines and ponderomotive potential iso-surface corresponding to the cusp-line example introduced in Eq.~\ref{eq:p_cusp} for $\alpha=1$. 
}
\end{figure}

Using the harmonic extension introduced above, we can introduce a broad class of coplanar trap networks as follows:
For a given analytic generator $P(x,y)$, we require that $\Phi(x,y,0)=0$ (so that $E_x$ and $E_y$ vanishes on $z=0$) and that $\partial_z \Phi(x,y,z)=P(x,y)$ so that $E_z$ and consequently $\bm{E}$ and $U_{\rm PP}$ vanishes in $z=0$ exactly at the zeros of $P$.

More explicitly, for any analytic generator $P(x,y)$ we define
\begin{align}
\label{eq:phip}
\Phi_P(x,y,z)&=\Phi_{\rm odd}[P](x,y,z)\\
&= \sum_{m=0}^{\infty}\frac{(-1)^m z^{2m+1}}{(2m+1)!}\Delta_{xy}^m P(x,y).
\end{align}
Then $\nabla^2\Phi_P=0$ and the planar null set for $\nabla \Phi$ is $P=0$.

As a minimal non-smooth example we consider 
\begin{align}
\label{eq:p_cusp}
P_{\rm cusp}(x,y)=y^2-\alpha^2 x^3,
\end{align}
whose zero set is parameterized by the cusp $(x,y)=(t^2,\alpha t^3)$ as illustrated in Fig.~\ref{fig:cusp}.
Because $\Delta_{xy}P_{\rm cusp}=2-6\alpha^2 x$, the series truncates:
\begin{align}
\Phi_{\rm cusp}=z(y^2-\alpha^2 x^3)+z^3\left(\alpha^2 x-\frac13\right).
\end{align}

\section{Periodic trap networks}

\begin{figure}[t]
\centering
\includegraphics[width=1.0\linewidth]{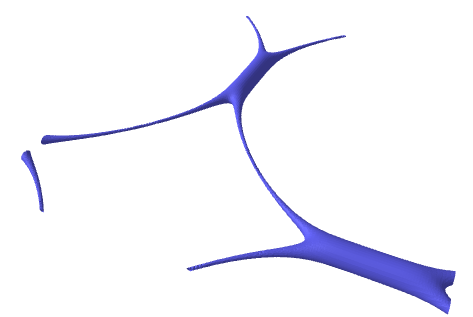}
\includegraphics[width=0.7\linewidth]{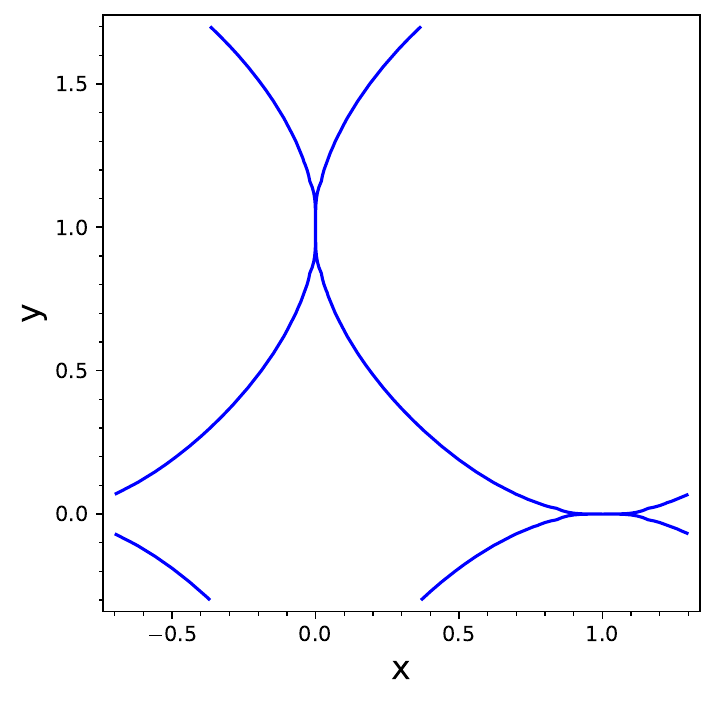}
\caption{
\label{fig:periodic}%
Null lines and ponderomotive potential iso-surface corresponding to the periodic network introduced in Eq.~\ref{eq:p_round} for $c=1/4$. 
}
\end{figure}


The odd harmonic extension is also well suited to periodic planar networks.
Let $P(x,y)$ be periodic with periods $L_x$ and $L_y$ and Fourier expansion
\begin{align}
P(x,y)=\sum_{\bm{k}} p_{\bm{k}}\exp\!\left[i\left(k_x x+k_y y\right)\right],
\end{align}
where $k_x=2\pi m/L_x$ and $k_y=2\pi n/L_y$ for $m,n \in \mathbb Z$.

Applying the odd extension mode by mode gives
\begin{multline}
\Phi_P(x,y,z)=p_{00}\,z\\
+\sum_{\bm{k}\neq(0,0)} p_{\bm{k}} \frac{\sinh\!\left(k \,z\right)} {k} \exp\!\left[i\left(k_x x+k_y y\right)\right].
\end{multline}
This expression satisfies $\nabla^2\Phi_P=0$, $\Phi_P(x,y,0)=0$, and $\partial_z\Phi_P(x,y,0)=P(x,y)$.
Thus the in-plane RF null set remains the zero contour $P(x,y)=0$.


As an example, we consider a family of rounded squares with intersections at the corners, providing a fully connected rectangular grid of traps as illustrated in Fig.~\ref{fig:periodic}.
\begin{multline}
\label{eq:p_round}
P_{\rm round}(x,y)=\cos(\pi x)\\
+\cos(\pi y)+c\left[(\cos(\pi x)-\cos(\pi y))^2-4\right].
\end{multline}
Expanding in Fourier modes gives
\begin{multline}
P_{\rm round}=-3c+\cos(\pi x)+\cos(\pi y)\\
+\frac{c}{2}\cos(2\pi x)+\frac{c}{2}\cos(2\pi y)\\
-c\cos(\pi(x+y))-c\cos(\pi(x-y)).
\end{multline}
Near an edge midpoint such as $(1,0)$, writing $x=1+u$ and $y=v$ gives the quadratic part
\begin{align}
P_{\rm round}^{(2)}=\pi^2\left[\left(\frac12-2c\right)u^2-\left(\frac12+2c\right)v^2\right].
\end{align}
Therefore the node remains a connected crossing for $|c|<1/4$ and becomes locally isolated for $|c|>1/4$.
We note that the structure of the intersection is the hexapole crossing identified in \cite{wesenberg2009:ideal_intersections_radiofrequency}, and provides no first-order confining force at the intersection point.
This example demonstrates that fully confining field-free trap networks are possible.

\section{Discussion and outlook}

In this work, we have only considered field synthesis not electrode geometry.
The odd extension reproduces the familiar four-wire linear guide when $P=x$:
This type of guide is challenging for surface-electrode traps, and it would be an obvious next step to include the even part of the harmonic extension to find field designs better suited for this technology.

Going further away from the odd extension, we may treat the full harmonic extension as merely a parametrization of the space of harmonic functions, and use it to create algebraic constraints for field characteristics on curves that are not in the plane. 
We have used this approach to identify field configurations for non-planar four-way cusp intersections with guide lines on the union of $(x,y,z)=(t^3,0,t^2)$ and $(x,y,z)=(0, t^3,t^2)$.


\bibliography{2026cuspfields}

\end{document}